\documentclass[prl,twocolumn,showpacs]{revtex4}
\usepackage{amssymb, amsmath}
\usepackage[dvips]{graphicx}
\usepackage{latexsym}
\usepackage{graphicx}
\usepackage{bm}
\newcommand{\be}{\begin{equation}}
\newcommand{\ee}{\end{equation}}
\newcommand{\ber}{\begin{eqnarray}}
\newcommand{\eer}{\end{eqnarray}}
\newcommand{\de}{\end{equation*}}
\newcommand{\cer}{\begin{eqnarray*}}
\newcommand{\der}{\end{eqnarray*}}
\begin{document}
\title{Fibre coupled dual-mode waveguide interferometer with $\lambda $/130 fringe
spacing}
\author{Richard
 M. Jenkins\footnote{Electronic address: rmjenkins@qinetiq.com} and Andrew F. Blockley}
\affiliation{Optical Research and Consulting Business Group,
QinetiQ, Malvern, Worcs WR14 3PS, U. K.}
\author{J.
Banerji\footnote{Electronic address: jay@prl.res.in}}
\affiliation{Theoretical Physics Division, Physical Research
Laboratory, Navrangpura, Ahmedabad 380 009, India}
\author{Alan R. Davies\footnote{Electronic address: alan@cs.rhul.ac.uk}}
\affiliation{Dept. of Computer Science, Royal Holloway, University
of London, Egham, Surrey TW20 0EX, U.K.}
\begin{abstract}
Predictions and measurements of a multimode waveguide interferometer
operating in a fibre coupled, ``dual-mode'' regime are reported.
With a 1.32 \ensuremath{\mu}m source, a complete switching cycle of
the output beam is produced by a 10.0 nm incremental change in the
8.0 \ensuremath{\mu}m width of the hollow planar mirror waveguide.
This equates to a fringe spacing of $\sim\lambda /130$. This is an
order of magnitude smaller than previously reported results for this
form of interferometer.
\end{abstract}
\maketitle

The propagation of light through rectangular and planar multimode
waveguides can result in interesting self-imaging effects based on
multimode interference (MMI) phenomena~\cite{1,2}. Over the last
decade or so, these effects have been demonstrated as the basis of
splitters~\cite{3,4,5}, modulators and switches ~\cite{6},
Mach-Zehnder interferometers~\cite{7} and laser resonators~\cite{8}.

More recently, Ovchinnikov and Pfau~\cite{9} (see also
reference~\cite{10}) have described a novel form of multimode
waveguide interferometer based on a planar waveguide formed from a
pair of fully reflecting mirrors. Light enters the planar waveguide
at some angle $+\theta$ exciting a spectrum of modes. The ensuing
multimode propagation and interference result in oscillations and
revivals in the transverse momentum of the propagating
field~\cite{9}. If $L$ is the length of the multimode guide,
$\lambda$ is the wavelength of the injected radiation, and $m$ is an
integer number, then a guide width $w=\sqrt{L\lambda/(4m)}$
maximises the magnitude of the momentum oscillations at the guide
exit. Under this condition, small changes in guide width cause the
output beam to swing back and forth between the angles of $\pm
\theta $.

In their experiment, Ovchinnikov and Pfau~\cite{9} coupled a 2.0 mm
diameter beam from a 0.633 \ensuremath{\mu}m source into a 50.0 mm
long planar waveguide at an angle of 0.25 radian. With a waveguide
width of 30 \ensuremath{\mu}m,  a complete cycle of the angular
deviation of the output beam was produced by a 70.0 nm change in
guide width. The sensitivity to the change in guide width equated to
a fringe spacing of $\lambda /9$~\cite{9}. This is substantially
smaller than that achieved with a Michelson interferometer.

Although this result is impressive, Ovchinnikov and Pfau~\cite{9}
suggested that narrower guide widths ($w\sim \lambda$) should lead
to improved sensitivity. They also alluded to the interesting case
where only two modes are excited in the planar multimode waveguide,
suggesting that the resulting fringe characteristic should be
uniform and periodic. If this situation could be realized in
practice,  then the interferometer could be used over a broader
range of contiguous mirror spacings as there would be no "collapse"
phenomena to impair operation. Furthermore, with the excitation of
only the two lowest order modes, the attenuation in the planar
mirror waveguide would be minimized. This is important as, what
ultimately limits the sensitivity of this form of interferometer,
are the larger attenuation coefficients of the higher order modes.

\begin{figure}[htbp]
\begin{center}
\includegraphics[width=6cm]{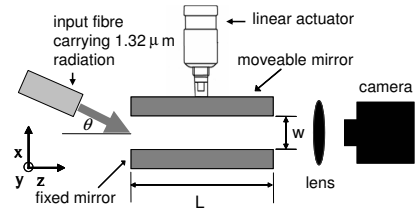}
\caption{ Schematic of the fibre coupled waveguide interferometer
illustrating the input beam and the camera configuration for
measuring the near-field output.\label{fig1}}
\end{center}
\end{figure}

\begin{figure}[htbp]
\begin{center}
\includegraphics[width=5.0cm]{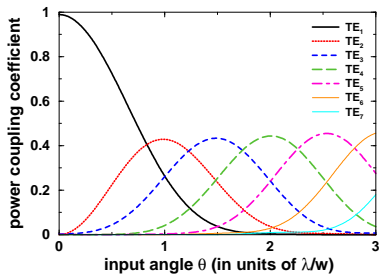}
\caption{ (color online). Overlap integral calculations illustrating
the power coupling coefficients for the modes TE$_p$ as a function
of angular misalignment for the case of a TEM$_{00}$ input
field.\label{fig2}}
\end{center}
\end{figure}

In this Letter, we show how to implement the proposed dual mode
operational regime. Our experimental configuration (see
Fig.~\ref{fig1}) was similar to that used by Ovchinnikov and
Pfau~\cite{9} .
The planar multimode waveguide was formed by two 50.0 mm diameter
fully reflecting gold coated mirrors both having a surface figure of
\ensuremath{\lambda}/10 at 632.8 nm. One of the mirrors was held in
a fixed precision mount, the other was mounted on a linear actuator.
In our case, 1.32 \ensuremath{\mu}m radiation from a Nd:YAG laser
source was coupled to the planar waveguide from a single-mode,
polarization maintaining fibre with an effective 1/e$^{2}$
TEM$_{00}$ mode diameter of 6.5 \ensuremath{\mu}m.  The fibre was
held straight in a fibre guide and adjusted so that the polarization
orientation of the output field was parallel to the plane of the
mirror surfaces, i.e. parallel to the y-axis in Fig.~\ref{fig1}.
Initially the fibre axis was aligned to be co-linear with the planar
waveguide axis and butted up to it. Under this condition a choice of
$w= 6.5/0.703 = 9.25$ \ensuremath{\mu}m, maximizes the power
coupling to its fundamental mode~\cite{11}. From this starting
point, the fibre was tilted by an angle  $\theta$ with respect to
the axis of the planar waveguide. In practice, although the last
centimeter of the fibre was stripped back to its 125
\ensuremath{\mu}m cladding diameter, this still meant that the axis
of the fibre pivoted about a 62.5 \ensuremath{\mu}m radius.
Depending on the magnitude of $\theta$, this leads to a short
free-space propagation distance and some diffraction before the beam
enters the planar waveguide. This was taken into account in the
overlap integral calculations to obtain the power coupling
coefficients as a function of the input angle $\theta$ (see Fig.
~\ref{fig2}).

As $\theta$ increases, modes of higher order are excited in turn.
The peak in the excitation characteristic of any given higher order
mode occurs when the angle of incidence of the input field
corresponds to the angle of one of the plane wave components of the
higher order mode itself. To a good approximation, to maximize the
excitation of the TE$_p$-th mode, an input beam angle of $\theta=\pm
p\lambda/(2w)$ is required. Hence for $p=2$, one obtains $\theta=\pm
\lambda/w=1.32/9.25=0.14$ radian. As indicated in Fig.~\ref{fig2},
by choosing an input angle of half this magnitude, i.e.
$\lambda/(2w)=0.07$ radian, the excitation is essentially limited to
two modes, TE$_1$ and TE$_2$.

Under this condition, we calculate the sensitivity of the output
field to changes in guide width. The phase change between any two TE
modes following propagation through an axial distance $L$, is given
by: $\phi_{pq} =L(\beta _p -\beta _q )$ where, $\beta_{p}$ is the
phase coefficient of the TE$_{p}$ mode: \be \beta _p
=\frac{2\pi}{\lambda}\left\{ {1-\left[ {\frac{p\lambda }{2w}}
\right]^2} \right\}^{1/2}. \label{two}\ee Under the condition
$(p\lambda /2w)^{2}<<1$, we  get \be \phi _{pq } =L\frac{\pi \lambda
}{4w^2}(q^2-p^2).\label{three}\ee Putting $p = 1$ and $q= 2$ for the
modes TE$_{1}$ and TE$_{2}$ respectively, differentiating with
respect to the guide width and rearranging, yields: \be
\partial \phi _{12 } =-\frac{3}{2}\frac{L\pi \lambda
}{w^3}\partial w.\label{four}\ee Equating Eq.~\ref{four} to $2\pi $
gives the incremental change in guide width that will produce a
$2\pi $ phase change between the modes, and hence, a change in the
output beam angle from $+\lambda /(2w)$ to $-\lambda /(2w)$, and
back again, i.e. a fringe, as \be
\partial w=-\frac{4}{3}\frac{w^3}{L\lambda
}.\label{five}\ee Eq.~\ref{five} leads us to conclude that {\it
small guide widths in conjunction with long waveguides and long
wavelength radiation produce maximum sensitivity}. However, reducing
the guide width and increasing both the guide length and the
wavelength also cause the attenuation of the excited modes to
increase in a differential manner. As shown below, this will affect
the power ratio between the modes and impact on our ability to
measure variations in the output field due to incremental changes in
guide width.

The fractional power transmission for the mode TE$_{p}$ through a
planar waveguide of length $L$ is given by \be t_p=\exp(-2\alpha _p
L),\quad\alpha _p =\frac{\lambda ^2p^2}{2w^3}\rm{Re}[(
\epsilon^2-1)^{-1/2}]\label{seven}\ee Here, $\alpha _p$ is the
attenuation coefficient and $\epsilon=n -ik$ is the complex
refractive index of the wall material.  For a 1.32 \ensuremath{\mu}m
source in conjunction with a 50 mm long planar waveguide formed from
gold ($n = 0.419$ and $k = 8.42$) coated mirrors, Eq.~\ref{seven}
yields  $t_p=\exp(-503p^2/w^3)$. For $w=7$ \ensuremath{\mu}m, this
yields fractional transmission values for the modes TE$_1$, TE$_2$
and TE$_3$ of 0.23, 0.0028 and 1.85$\times 10^{-6}$ respectively,
with $t_1/t_2\sim 81$. For $w=11$ \ensuremath{\mu}m, the
corresponding values are 0.68, 0.22 and 0.03, with $t_1/t_2\sim
3.1$. From this perspective, with the aim of working with small
guide widths in order to achieve high sensitivity, we refer back to
Fig.~\ref{fig2} and opt for an input angle of $\pm \lambda /w$. This
provides the highest starting magnitude of TE$_2$, while the
additional excitation of the modes TE$_3$ and TE$_4$ are of little
consequence because of their significantly higher attenuation.

\begin{figure}[htbp]
\begin{center}
\includegraphics[width=6cm]{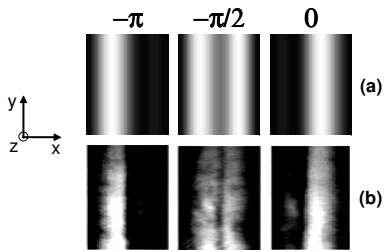}
\caption{ Near field intensity profiles as a function of the phase
difference between the modes TE$_{1}$ and TE$_{2}$: (a) predictions;
(b) measured profiles.\label{fig3}}
\end{center}
\end{figure}

On the aforementioned basis, we started off with a guide width of
11.0 \ensuremath{\mu}m with the aim of  making measurements of
interferometer sensitivity as the width was decreased to 8
\ensuremath{\mu}m. The launch angle $\lambda /w$ was kept fixed at
0.14 radian corresponding to a median guide width of 9.25
\ensuremath{\mu}m.  To start with, the mirrors were very accurately
aligned with respect to one another to ensure that they were
parallel. To aid in this process and confirm that dual mode
operation was achieved in practice, a magnified (~100 times) image
of the field generated at the exit of the multimode waveguide
interferometer was produced with a microscope objective and viewed
with a Hamamatsu infrared vidicon camera C2400-03. The latter has an
operational waveband of 0.8 - 2.1 \ensuremath{\mu}m and a resolution
of 720 $\times$ 576 pixels. For a guide width of 11
\ensuremath{\mu}m and an input angle of 0.14 radian, our predicted
transverse intensity profiles at the exit of the planar waveguide
are shown in Fig.~\ref{fig3}a. These correspond to a
TE$_{1}$:TE$_{2}$ output power ratio of $1.86:1$ and phase
differences of $-\pi$, $-\pi/2$ and $0$ radians induced between
these modes by the displacement of the moveable mirror.
Fig.~\ref{fig3}b shows the results of equivalent measurements made
with the camera. The widths of the images correspond to the starting
multimode guide width of 11 \ensuremath{\mu}m while the heights
correspond to ~100 \ensuremath{\mu}m high segments of the total
vertical extent of the output fields. As a consequence of
magnification, the physical dimensions of the images in
Fig.~\ref{fig3}b are ~ 1.1 x 10 mm. For presentational purposes, the
heights of the images have been compressed by a factor of three. The
very good agreement between the measured and predicted field
intensity contours confirms that dual mode operation was achieved in
practice.

\begin{figure}[htbp]
\begin{center}
\includegraphics[width=6.5cm]{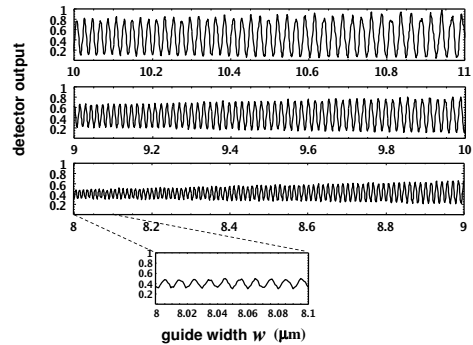}
\caption{ Experimental measurements of the fringe characteristics as
a function of varying guide width \textit{w} from 11 to 8
\ensuremath{\mu}m. The characteristic illustrates uniformity and
periodicity over many cycles with the absence of revival and
collapse phenomena in conjunction with decreasing fringe spacing
with decreasing guide width. The bottom figure is a higher
resolution plot indicating a fringe spacing of \ensuremath{\sim} 10
nm with a guide width of \ensuremath{\sim} 8
\ensuremath{\mu}m.\label{fignew}}
\end{center}
\end{figure}

In order to demonstrate that dual mode operation was achieved over a
wide range of contiguous mirror spacings without the presence of the
``collapse'' phenomena associated with multimode operation, we made
a further measurement. This involved accurately locating an InGaAs
photo-detector with an effective diameter of 1.0 mm in the plane of
the magnified image of the output field. The detector was offset
from the axis of the planar multimode waveguide such that its active
area only captured light from one side of the image. A linear
voltage ramp was then applied in an incremental manner to our
Newport ESA 1330 electro-strictive actuator. At the same time the
output from the photo-detector was digitized and recorded. The
starting value, and the magnitude of the applied voltage ramp, was
chosen to correspond to the most linear portion of the displacement
versus applied voltage characteristic of the actuator and to change
the planar guide width from an initial value of 11 \ensuremath{\mu}m
to a final value of 8 \ensuremath{\mu}m. The tolerance on the
resulting guide width was estimated to be $\pm$ 0.5
\ensuremath{\mu}m.   A plot of the detector output amplitude as a
function of the guide width is illustrated in Fig.~\ref{fignew}.
\vskip 0.5cm
A continuous fringe characteristic is observed. Despite decreasing
visibility, it exists over many periods ($>100$) and exhibits no
collapse phenomena.  The bottom figure of the experimental data set
illustrates a higher resolution plot indicating a fringe spacing of
\ensuremath{\sim} 10 nm with a guide width of \ensuremath{\sim} 8
\ensuremath{\mu}m. The decreasing visibility of the fringes with
decreasing guide width is a consequence of  increased aperturing at
the entrance plane, the larger attenuation of both modes, and the
higher relative attenuation of the TE$_{2}$ mode compared with the
fundamental. For values of guide width much below 8.0
\ensuremath{\mu}m the TE$_{2}$ mode amplitude is so small that the
fringe visibility becomes comparable with concatenated measurement
noise.

In order to more accurately confirm the sensitivity of the
interferometer, a further measurement was undertaken. This involved
making use of the fringe characteristic itself to calibrate the
change in guide width produced by the actuator against a high
precision micrometer which was incorporated on the same
translational stage. With the actuator calibrated in this manner the
applied voltage was changed manually in order to produce ten
complete fringe cycles (of the form shown in Fig.~\ref{fig3}b) as
observed directly on the vidicon camera with the naked eye. Using
this approach, the error in the measurement of the incremental
change in guide width necessary to produce a complete switching
cycle of the output field was estimated to be of the order of $\pm$
2.0 nm

\begin{figure}[hptb]
\begin{center}
\includegraphics[width=5.0cm]{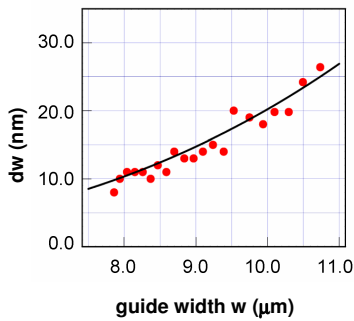}
\caption{ Experimental measurements (points) and theoretical
predictions (solid line based on Eq.~\ref{five}) of the incremental
change in guide width required to produce a complete switching cycle
of the near-field output pattern (as illustrated in
Figure~\ref{fig3}) as a function of guide width.\label{fig5}}
\end{center}
\end{figure}

As illustrated in Fig.~\ref{fig5}, these results were plotted as a
function of absolute guide width (measured to an estimated accuracy
of   $\pm$ 0.5 \ensuremath{\mu}m) together with a theoretical
prediction based on Eq.~\ref{five}.
With a guide width of 8.0 \textit{\ensuremath{\mu}}m, the measured
incremental change in guide width required to produce a complete
switching cycle of the output beam was \ensuremath{\sim} 10.0 nm in
agreement with the earlier result. For our 1.32
\textit{\ensuremath{\mu}}m source, this equates to a change in guide
width of \ensuremath{\sim} \textit{\ensuremath{\lambda}}/130.

In conclusion, an improved multimode waveguide interferometer of the
form originally conceived by Ovchinnikov and Pfau~\cite{9} has been
demonstrated. Our implementation differs from previous work~\cite{9}
in three main ways: (i) a fibre is used to more efficiently couple
the light from the laser source into the planar waveguide, thereby
significantly reducing aperturing losses; (ii) the choice of the
TEM$_{00}$ waist diameter and the input angle, coupled with the
differential attenuation of the excited modes, ensures that only the
two lowest order modes (TE$_{1}$ and TE$_{2}$) are present at the
exit plane of the planar waveguide. This leads to the lowest
possible attenuation in the waveguide and circumvents the problems
of revival and collapse phenomena associated with the excitation of
more than two modes~\cite{9}. Finally, (iii) the changes in the
transverse mode output from the interferometer are measured in the
near field. With this implementation a complete cycle of the output
field pattern was produced by an incremental change in guide width
of \ensuremath{\sim}10.0 nm or \ensuremath{\sim}
\ensuremath{\lambda}/130. This should be compared with the
\ensuremath{\lambda}/9 change in guide width required to produce a
full switching cycle demonstrated in earlier work~\cite{9}.

A simple analytical expression, given by Eq.~\ref{five}, has been
derived for the sensitivity of the dual-mode interferometer and the
way it scales with guide width, guide length and wavelength. Using
Eq.~\ref{five}, predictions of the incremental change in guide width
required to produce a complete switching cycle of the output beam
are found to be in good agreement with the experimentally measured
results as illustrated in Fig.~\ref{fig5}.

With respect to achieving higher sensitivity, because of the
$\lambda^2$ dependence of the attenuation coefficient in Eq.
~\ref{seven}, we conclude that it is best to use the shortest
operational wavelength possible and, in relation to Eq. ~\ref{five},
to compensate for this by using longer guides and smaller guide
widths.  Further reduction in attenuation (and hence improved
sensitivity) might also be achieved with multilayer mirror coatings
designed to provide very high reflectivity for grazing angle
incidence TE fields.

We end by noting that dual-mode excitation  could also be produced
by a laterally offset TEM$_{00}$ beam or by a suitably offset
fundamental mode field from an input waveguide. As in the approach
described herein, this would result in an output field whose
intensity maximum switches from one side of the guide exit to the
other. This could be measured with a single or dual-element detector
placed directly at the guide exit.  This arrangement would make the
interferometer more compact facilitating its realisation in silicon
based MEMS technology. Thus, multimode waveguide interferometers of
the type originally proposed by Ovchinnikov and Pfau~\cite{9} should
find many sensor and switching applications in the more efficient
and more sensitive fibre-coupled dual-mode regime, demonstrated in
this paper.

\end{document}